\documentclass[9pt]{entcs}
\usepackage{entcsmacro}
\usepackage{graphicx}

\usepackage{amsmath}
\usepackage{amssymb}
\usepackage{float} 

\newcommand{\refeq}[1]{(\ref{#1})}

\def\lastname{Bianchetti, Marenco}
\begin{document}
\begin{frontmatter}
    \title{Valid inequalities and a branch-and-cut algorithm for the routing and spectrum allocation problem} \author{Marcelo Bianchetti\thanksref{UBAAddress}\thanksref{UNGSAddress}\thanksref{mbemail}}, 
    \author{Javier Marenco\thanksref{UBAAddress}\thanksref{UNGSAddress}\thanksref{jmemail}}
    \address[UBAAddress]{Departamento de Computaci\'on, FCEyN \\ Universidad de Buenos Aires\\
    Buenos Aires, Argentina}
    \address[UNGSAddress]{Instituto de Ciencias\\ Universidad Nacional de General Sarmiento\\
    Buenos Aires, Argentina}
    \thanks[mbemail]{Email:
    \href{mailto:mbianchetti@dc.uba.ar} {\texttt{\normalshape
        mbianchetti@dc.uba.ar}}}
    \thanks[jmemail]{Email:
    \href{mailto:jmarenco@dc.uba.ar} {\texttt{\normalshape
        jmarenco@dc.uba.ar}}}
\begin{abstract} 
    One of the most promising solutions to deal with huge data traffic demands in large communication networks is given by flexible optical networking, in particular the \emph{flexible grid} (flexgrid) technology specified in the ITU-T standard G.694.1. In this specification, the frequency spectrum of an optical fiber link is divided into narrow frequency \emph{slots}. Any sequence of consecutive slots can be used as a simple channel, and such a channel can be switched in the network nodes to create a \emph{lightpath}.
    In this kind of networks, the problem of establishing lightpaths for a set of end-to-end demands that compete for spectrum resources is called the \emph{routing and spectrum allocation problem} (RSA). Due to its relevance, RSA has been intensively studied in the last years. \textcolor{red}{It has been shown to be NP-hard} and different solution approaches have been proposed for this problem. In this paper we present several families of valid inequalities, valid equations, and optimality cuts for a natural integer programming formulation of RSA and, based on these results, we develop a branch-and-cut algorithm for this problem. Our computational experiments suggest that such an approach is effective at tackling this problem.
\end{abstract}
\end{frontmatter}
\section{Introduction}\label{intro}
    \emph{Flexgrid optical networks} are an emerging technology in the field of optical networks. In these networks, the frequency spectrum is divided into narrow frequency \emph{slots}, and a sequence of consecutive slots forms a \emph{channel} that can be switched in the network nodes to create a \emph{lightpath} between two nodes. The \emph{routing and spectrum allocation problem} (RSA) \cite{undirectedRSA,Slice_ITU,directedRSA} consists in establishing the lightpaths for a set of end-to-end traffic demands that are expressed in terms of the number of required slots.

Formally, we are given a digraph $G=(V,E)$ representing the optical fiber network, a fixed number $\bar s\in\mathbb{Z}_+$ of available slots, and a set of \emph{demands} $D=\{d_i = (s_i,t_i,v_i)\}_{i=1}^k$, where each demand $d_i$, $i=1,\dots,k$, is composed by a source $s_i\in V$, a target $t_i\in V$, and a volume $v_i\in\mathbb{Z}_+$. We define a \emph{lightpath} for a demand $d_i=(s_i,t_i,v_i)$ to be a tuple $(l,r,p)$, where $1\le l\le l+v_i-1\le r \le \bar s$ and $p$ is a (directed) path in $G$ from $s_i$ to $t_i$. In this setting, \textcolor{red}{RSA consists in establishing a lightpath associated to each demand}, in such a way that lightpaths do not overlap. In other words, each demand $d_i$, for $i=1,\dots,k$, must be assigned a path $p_i$ (regarded as a sequence of arcs) in $G$ between $s_i$ and $t_i$ and an interval $[l_i,r_i]$ consisting of at least $v_i$ consecutive slots in $[1,\bar s]$ in such a way that if $p_i\cap p_j \ne \emptyset$ then $[l_i,r_i] \cap [l_j,r_j] = \emptyset$, for any two demand indices $i\ne j$. Deciding if RSA is feasible for a particular graph $G$, a set of demands $D$, and an amount of slots $\bar s$ is NP-complete \cite{RSA_NP}, so calculating the minimum $\bar s$ such that RSA is feasible is NP-hard. Furthermore, these problems turn out to be quite difficult to solve in practice.

\subsection{Integer programming formulation}

In \cite{RSAmodels_springer} several integer programming models for RSA are presented. In this work, we concentrate on the so-called \emph{DSL-BF formulation}, which was one of the best-performing models in the experiments reported in \cite{RSAmodels_springer}. For every demand $d \in D$, every arc $e \in E$, and every slot $s \in S = \{1, \dots , \bar s\}$, we introduce the binary variable $u_{des}$, in such a way that $u_{des} = 1$ if and only if the demand $d$ uses the slot $s$ over the arc $e$. For technical reasons, we also consider the fictitious variable $u_{de,\bar s+1}$ that always takes value 0, for every $d \in D$ and every $e \in E$.

If $d = d_i = (s_i,t_i,v_i) \in D$ is a demand, for some $i\in\{1,\dots,k\}$, we define $s(d) = s_i$, $t(d) = t_i$, and $v(d) = d_i$. For $j\in V$, we define $\delta^-(j)$ to be the set of incoming arcs to $j$, and $\delta^+(j)$ to be the set of outgoing arcs from $j$. We also define $\delta(j) = \delta^-(j) \cup \delta^+(j)$. In this setting, the \emph{DSL-BF formulation} for RSA is the following integer program.
\begin{alignat}{4}
    \text{min} && \sum_{d\in D} \sum_{e\in E} \sum_{s \in S} u_{des}/v(d) \quad && &\label{DSL_BF:OF}\\
    \mbox{s.t.} && \quad\sum_{e \in \delta^-(j)} u_{des} - \sum_{e \in \delta^+(j)} u_{des} &= 0 &
        \begin{tabular}{r}
            $\forall d \in D$, 
            $\forall j \in V\backslash\{s(d),t(d)\}$, 
            $\forall s \in S$
        \end{tabular}& \label{DSL_BF:2}\\
    && \sum_{e \in \delta^+(s(d))} \sum_{s\in S} u_{des} &\geq v(d) &
    \begin{tabular}{r}
        $\forall d \in D$
    \end{tabular}& \label{DSL_BF:4}\\
    && \sum_{e \in \delta^-(s(d))} \sum_{s \in S} u_{des} &= 0 &
    \begin{tabular}{r}
      $\forall d \in D$
    \end{tabular}& \label{DSL_BF:3}\\
    && \sum_{d \in D} u_{des} &\leq 1 &
    \begin{tabular}{r}
        $\forall e \in E$, $\forall s \in S$
    \end{tabular}&\label{DSL_BF:5}\\
    && v(d)(u_{des} - u_{de, s+1}) &\leq \sum_{s' = f}^s u_{des'} &\quad\quad
    \begin{tabular}{r}
        $\forall d \in D$, $\forall e \in E$,
        $\forall s \in S$,
        $f = \mbox{max}\{1, s-v(d)+1\}$
    \end{tabular}&\label{DSL_BF:7} \\
    && u_{des} &\in \{0,1\} & 
    \begin{tabular}{r}
        $\forall d \in D$, $\forall e \in E$,
        $\forall s \in S$
    \end{tabular}&\label{DSL_BF:8}
\end{alignat}

The objective function \refeq{DSL_BF:OF} asks for the total length of the lightpaths to be minimized. This also forbids spurious cycles in any optimal solution. Constraints \refeq{DSL_BF:2} impose flow conservation restrictions for each demand at each node in the network, except for the source and sink nodes associated with the demand. Constraints \refeq{DSL_BF:4} ensure that the required number of slots is routed for each demand, \textcolor{red}{whereas constraints \refeq{DSL_BF:3} forbid incoming arcs into the source node of each demand}. Constraints \refeq{DSL_BF:5} guarantee that no two different lightpaths overlap. Finally, constraints \refeq{DSL_BF:7} ensure slot contiguity: if $u_{des} - u_{de, s+1} > 0$ then $s$ is the last slot used by the demand $d$ on the arc $e$, and in this case we ask all variables $u_{dei}$ for $i=s-v(d)+1,\dots,s$ to be activated. These constraints use the fact that the fictitious variable $u_{de,\bar s+1}$ takes value 0, for $d\in D$ and $e\in E$.

We define $RSA(G,D,\bar s)$ to be the convex hull of feasible solutions of the DSL-BF formulation \refeq{DSL_BF:2}-\refeq{DSL_BF:8}. A \emph{valid inequality} (resp.~\emph{valid equation}) is a linear inequality (resp.~equation) on $u$ satisfied by all points in $RSA(G,D,\bar s)$. An \emph{optimality cut} is a non-valid inequality that does not remove all optimal solutions. The symmetries present in the DSL-BF formulation give rise to the following technical results.
\begin{theorem} \label{thm1}
If $a\cdot u \leq b$ is a valid inequality (resp., optimality cut) for $RSA(G,D,\bar s)$, then the \emph{slot-symmetrical inequality}
\begin{equation}
  \sum_{d \in D}\sum_{e \in E}\sum_{s \in S} a_{des}~ u_{de,\bar s-s+1} \leq b \label{vi:slot-symmetrical:1}
\end{equation}
is also valid 
(resp., an optimality cut) 
for $RSA(G,D,\bar s)$. Furthermore, if $a\cdot u \leq b$ induces a facet of $RSA(G,D,\bar s)$, then \refeq{vi:slot-symmetrical:1} also induces a facet of this polytope.
\end{theorem}

\begin{theorem} \label{thm2}
Let $j\in V$ such that $j\not\in\{s(d), t(d)\}$ for every $d\in D$. Let $a\cdot u \le b$ be a valid inequality (resp., optimality cut) for $RSA(G,D,\bar s)$ such that for every $d\in D$ and every $s\in S$ there exist $\alpha_{ds}\in\mathbb{R}$ and $\beta_{ds}\in\mathbb{R}$ with $a_{des} = \alpha_{ds}$ for every $e\in\delta^+(j)$ and $a_{des} = \beta_{ds}$ for every $e\in\delta^-(j)$. Then, the inequality
\begin{equation}
\sum_{s\in S} \sum_{d\in D} \left(\sum_{e\in\delta^-(j)} \alpha_{ds} u_{des} + \sum_{e\in\delta^+(j)} \beta_{ds} u_{des} + \sum_{e\in E\backslash\delta(j)} a_{des} u_{des} \right) \le b\label{flow-symmetrical}
\end{equation}
is also valid (resp., an optimality cut) for $RSA(G,D,\bar s)$.
\end{theorem}

\begin{theorem} \label{thm3}
Fix $d\in D$ and define $\delta_d = \delta(t(d))\cup \delta(s(d))$. Let $a\cdot u \le b$ be a valid inequality (resp., optimality cut) for $RSA(G,D,\bar s)$ such that for every $s\in S$ there exist $\alpha_{ds}\in\mathbb{R}$ and $\beta_{ds}\in\mathbb{R}$ with $a_{des} = \alpha_{ds}$ for every $e\in\delta^+(s(d))$, $a_{des} = \beta_{ds}$ for every $e\in\delta^-(t(d))$, $a_{des}=0$ for every $e \in\delta^-(s(d))\cup\delta^+(t(d))$, and $a_{d'es} = 0$ for every $d'\ne d$ and $e\in\delta_d$. Then, the inequality
\begin{equation}
\sum_{s\in S} \left( \sum_{e\in\delta^-(t(d))} \alpha_{ds} u_{des} + \sum_{e\in\delta^+(s(d))} \beta_{ds} u_{des} + \sum_{d'\in D}\sum_{e\in E\backslash\delta_d} a_{d'es} u_{d'es} \right) \le b\label{src-symmetrical}
\end{equation}
is also valid (resp., an optimality cut) for $RSA(G,D,\bar s)$.
\end{theorem}

The polytope $RSA(G,D,\bar s)$ seems hard to study, since even characterizing its dimension is not straightforward (as suggested by the valid equations in the next sections). Due to this fact, we explore in this work families of valid inequalities without providing facetness results. Instead, we aim for the more modest goal of showing that --together with the model constraints-- the inequalities from these families do not imply each other.

    \section{Optimality cuts coming from flow considerations}

In this and the following sections we propose several families of valid inequalities, valid equations, and optimality cuts for $RSA(G,D,\bar s)$. In particular, in this section we only present optimality cuts. 

For each demand $d\in D$, we can eliminate cycles involving $t(d)$ with the analogous version of constraints \refeq{DSL_BF:3} applied to $\delta^+(t(d))$. We can also eliminate cycles involving intermediate nodes with the optimality cut
\begin{align}
  & \sum_{e \in \delta^+ (i)} u_{des} \leq 1 &
  \begin{tabular}{r}
    $\forall d \in D$, 
    $s \in S$,
    $\forall i \in V\setminus \{t(d)\}$.
  \end{tabular} \label{vi:oneSlotOnceFrom:1}
\end{align}
We call this family \emph{oneSlotOnceFromV}, and we call \emph{oneSlotOnceFromSrc} the subset obtained by taking $i = s(d)$.

The model constraints ask for at least $v(d)$ slots to be assigned to demand $d$. We can impose that the number of slots assigned to demand $d$ does not exceed this value, with the so-called \emph{exactlyVdFromV} inequality:
\begin{align}
  & \sum_{e \in \delta^+ (i)} \sum_{s\in S} u_{des} \leq v(d)&
    \begin{tabular}{r}
      $\forall d \in D$, $\forall i\in V$.
    \end{tabular} \label{vi:exactlyVdFromSrc:1}
\end{align}
This is not a valid inequality (since the model allows more than $v(d)$ slots assigned to $d$), but is an optimality cut instead, since there always exists an optimal solution with exactly $v(d)$ slots assigned to $d$. When $i=s(d)$, we refer to this family as \emph{exactlyVdFromSrc}.

The next optimality cut states that if a slot $s$ in an arc $e\in\delta^+(i)$ is used by $d$, then no other outgoing arc from $i$ can be used by $d$, namely
\begin{align}
  & \sum_{e' \in \delta^+ (i) \setminus\{e\}} \sum_{s'\in S} u_{de's'} \leq v(d) (1 - u_{des})&
    \begin{tabular}{r}
      $\forall d \in D$, $\forall i\in V$, $\forall e \in \delta^+ (i)$,
      $s \in S$.
    \end{tabular} \label{vi:notBranchFromSrc:1}
\end{align}
This inequality also forces $d$ to use at most $v(d)$ slots. We call this family \emph{notBranchFromV}, and we call \emph{notBranchFromSrc} the subset obtained by taking $i=s(d)$. None of these inequalities is implied by the model constraints \refeq{DSL_BF:2}-\refeq{DSL_BF:7} and the inequalities \refeq{vi:exactlyVdFromSrc:1}, or vice versa.

The inequality
\begin{align}
  & \sum_{e \in E} u_{des'} \leq 
  \sum_{e \in E} u_{des} + 
  |E| \Big(1 - \sum_{e \in \delta^+ (s(d))} u_{des} \Big) &
  \begin{tabular}{r}
    $\forall d \in D$, 
    $ s,s' \in S$, $s' \ne s$
  \end{tabular} \label{vi:eqAmountOfAsForEachUsedS:1}
\end{align}
is valid for the model given by \refeq{DSL_BF:2}-\refeq{DSL_BF:7} and \refeq{vi:oneSlotOnceFrom:1} with $i=s(d)$. This inequality says that if the slot $s$ is used by $d$ then for any other slot $s'$ the amount of arcs in which $s'$ is used by $d$ must be at most the amount of arcs in which $d$ uses $s$. We can sum these inequalities over all slots, thus generating another valid inequality, which says that for every demand $d$ if a slot $s$ is used by $d$ then the amount of slots used by the demand all along the graph must be $v(d)$ times the amount of arcs in which $s$ is used. Given the constraints \refeq{DSL_BF:2}-\refeq{DSL_BF:7} and the inequalities \refeq{vi:oneSlotOnceFrom:1} and \refeq{vi:exactlyVdFromSrc:1}, this new family is not implied by \refeq{vi:eqAmountOfAsForEachUsedS:1}.

Given an arc subset $E' \subseteq E$, we call $P_{E'}$ to the set of all paths in $E'$ without cycles, and we call $M(P_{E'}) \subseteq E'$ to a set of vertex-disjoint paths with maximum total number of arcs. Any optimal solution can use at most $|V(M(P_{E'}))|-|M(P_{E'})|$ arcs on $E'$. This results in the following large family of optimality cuts:
\begin{align}
  & \sum_{s \in S}\sum_{e \in E'} u_{des} \leq v(d) \Big( |V(M(P_{E'}))|-|M(P_{E'})| \Big)&
    \begin{tabular}{r}
      $\forall d \in D$,
      $\forall E' \subseteq E$.
    \end{tabular} \label{vi:maximumSetOfPaths:2}
\end{align}
Constraints \refeq{DSL_BF:2}-\refeq{DSL_BF:7} and \refeq{vi:maximumSetOfPaths:2} imply the family \emph{exactlyVdFromV}.

Given an arc $ij \in E$, we define $\delta^-(i) \cup \delta(j)$ to be the  \emph{incoming double broom} (InDBroom) associated with $ij$. Similarly, we define $\delta^+(i) \cup \delta(j)$ to be the \emph{outgoing double broom} (OutDBroom) associated with $ij$. If $DB$ is either an InDBroom or an OutDBroom in $G$, then $|M(P_{DB})| = 3$. The particular cases of \refeq{vi:maximumSetOfPaths:2} in which $E'$ is an InDBroom or an OutDBroom are called \emph{incomingDBrooms} and \emph{outcomingDBrooms}, respectively.

Inequalities \refeq{vi:inducedArcs:1} and \refeq{vi:inducedArcs:2} are optimality cuts for $RSA(G,D,\bar s)$. In particular, \refeq{vi:inducedArcs:2} is a subset of the inequalities \refeq{vi:maximumSetOfPaths:2}. Constraints \refeq{DSL_BF:2}-\refeq{DSL_BF:7} and \refeq{vi:inducedArcs:2} do not imply the family \emph{exactlyVdFromV}, nor vice versa. 
\begin{align}
  & \sum_{e \in E'} u_{des} \leq (|V(E')|-1)&
    \begin{tabular}{r}
      $\forall d \in D$, 
      $\forall s \in S$,
      $\forall E' \subseteq E$,
    \end{tabular} \label{vi:inducedArcs:1}\\
  & \sum_{s \in S}\sum_{e \in E'} u_{des} \leq v(d) (|V(E')|-1)&
    \begin{tabular}{r}
      $\forall d \in D$,
      $\forall E' \subseteq E$.
    \end{tabular} \label{vi:inducedArcs:2}
\end{align}
In our experiments, we consider the subfamilies of \refeq{vi:inducedArcs:1} and \refeq{vi:inducedArcs:2} generated by taking $E'$ to be (a) a cycle in $G$, (b) an undirected cycle in $G$ (i.e., disregarding the arc orientations), (c) the set of all edges induced by the vertices of a cycle in $G$, and (d) the set $\{ij, ji\}$, when both arcs exist between the nodes $i$ and $j$.

Finally, for two arcs $ij, ji\in E$ (when such a structure exists), we consider the inequality
\begin{align}
  & \sum_{s' \in S} u_{d,ji,s'} \leq v(d) (1 - u_{d,ij,s})&
    \begin{tabular}{r}
      $\forall ij,ji \in E$, $\forall d \in D$,
      $\forall s \in S$.
    \end{tabular} \label{vi:simplestCyclesImplication:1}
\end{align}
This inequality states that if the demand $d$ uses the slot $s$ in the arc $ij$, then it cannot use any slot in the arc $ji$, which clearly holds in any optimal solution. These inequalities are not implied by the model constraints \refeq{DSL_BF:2}-\refeq{DSL_BF:7} together with \refeq{vi:inducedArcs:1}, \refeq{vi:inducedArcs:2}, and \emph{exactlyVdFromV}.

    \section{Contiguity inequalities}

In this section we present valid inequalities and equations dealing with the contiguity requirements, namely that each demand must use consecutive slots. These families are based on a result of \cite{RSAmodels_springer} where given a demand $d$, the contiguity constraint is obtained mainly by grouping the slots that are distanced from each other by the volume of $d$. The first family is called \emph{contiguityIneqs}, and is defined by
\begin{align}
    &\sum_{
      \begin{subarray}{c} 
        s \in \{1, \dots, i\}:\\ 
        s ~ \equiv ~ i~(v(d)) 
      \end{subarray} } u_{des}
    ~ \geq ~
    \sum_{
      \begin{subarray}{c} 
        s \in \{1, \dots, i-1\}:\\ 
        s +1 ~ \equiv ~ i~(v(d)) 
      \end{subarray} } u_{des} &
    \begin{tabular}{r}
        $\forall i \in S$.
    \end{tabular}\label{vi:contiguityIneqs:1}
\end{align}
If we furthermore assume that each demand $d\in D$ uses exactly $v(d)$ slots (there is always an optimal solution satisfying this requirement), then the following equations are valid for $RSA(G,D,\bar s)$:
\begin{align}
  &\sum_{
  \begin{subarray}{c} 
    s \in S:\\ 
    s ~ \equiv ~ i~(v(d)) 
  \end{subarray} } u_{des}
  ~ = ~
  \sum_{
    \begin{subarray}{c} 
      s \in S:\\ 
      s +1 ~ \equiv ~ i~(v(d)) 
    \end{subarray} } u_{des} &
  \begin{tabular}{r}
      $\forall d\in D$, $\forall e \in E$,
      $\forall i \in \{1, \dots, v(d)\}$.
  \end{tabular}\label{vi:contiguityEqs:1}
\end{align}

If $s,t\in V$, $s\ne t$, we define $P(G,s,t)$ to be the set of all directed paths from $s$ to $t$ in $G$. We define a \emph{minimal (s,t)-cut} to be a subset $C\subseteq E$ such that for every $p\in P(G,s,t)$ there exists $e \in C$ with $e \in p$, but for every $e'\in C$, $e'\ne e$, we have $e'\not\in p$. For every demand $d\in D$ and every minimal $(s(d),t(d))$-cut $S_c$, the lightpath associated with $d$ in any optimal solution contains exactly one arc from $S_c$. Let $SC_d$ be the set of minimal $(s(d),t(d))$-cuts for $d$. Then, the following
\begin{align}
&\sum_{e\in C} \sum_{
  \begin{subarray}{c} 
    s \in S:\\ 
    s ~ \equiv ~ i~(v(d)) 
  \end{subarray} } u_{des} ~ = ~ 1 &
\begin{tabular}{r}
  $\forall d \in D$,
  $\forall i \in \{1, \dots, v(d)\}$,
  $C \in SC_d$
\end{tabular} \label{vi:ppalSlots:1}
\end{align}
are optimality cuts for $RSA(G,D,\bar s)$. The families \refeq{vi:contiguityIneqs:1}-\refeq{vi:ppalSlots:1} are sufficient to ensure that each demand $d\in D$ uses $0$ or exactly $v(d)$ consecutive slots in every arc $e \in E$ \cite{RSAmodels_springer}.

The special cases of \refeq{vi:ppalSlots:1} given by $C = \delta^+(s(d))$ and $C = \delta^-(t(d))$ are called \emph{ppalSlotsFromSrc} and \emph{ppalSlotsToDst} respectively. We also separate the sub-family obtained by taking only one sub-group of slots, for example by fixing $i=1$. Another special sub-family of \refeq{vi:ppalSlots:1} is obtained by restricting to demands with $2v(d) > \bar s$. In these cases, the $2v(d) - \bar s$ central slots of some arc in $\delta^+(s(d))$ must be used.

If a slot $s$ is used by a demand $d$ and $d$ uses exactly $v(d)$ slots, then $d$ should use slots located at distance at most $v(d)$ from $s$. If we define $S' = \{1, \dots, s-v(d)\} \cup \{s+v(d), \dots, \bar s\}$, we can force this fact with
\begin{align}
  & \sum_{s' \in S'} u_{des'}\leq M (1 - u_{des})&
    \begin{tabular}{r}
      $\forall e \in E$, $\forall d\in D$,
      $\forall s \in S$,
      $M = \min\{|S'|, v(d)\}$,
    \end{tabular} \label{vi:farSlotsOff:1}
\end{align}
which we call \emph{farSlotsOff}. These optimality cuts are not implied by the model constraints \refeq{DSL_BF:2}-\refeq{DSL_BF:7} together with \emph{exactlyVdFromV}, \refeq{vi:contiguityEqs:1}, and \refeq{vi:ppalSlots:1} with $C=\delta^+(s(d))$.

We also consider the inequalities obtained by applying Theorem~\ref{thm1} to the model constraints \refeq{DSL_BF:7}, which we call \emph{symmetricalBFContiguity}. These symmetrical inequalities are not implied by constraints \refeq{DSL_BF:7}. The constraints of the model \emph{DSL-ASCC} presented in \cite{RSAmodels_springer} are also valid optimality cuts that are not implied by the model constraints \refeq{DSL_BF:2}-\refeq{DSL_BF:7}, the flow inequalities \emph{exactlyVdFromV} and the inequalities \emph{symmetricalBFContiguity}. We call them \emph{contiguityASCC}.
    \section{Non-overlapping inequalities}

The contiguity and non-overlapping constraints imply that if a demand $d$ uses two slots $s_1$ and $s_3$ with $s_1 < s_3-1$, then no other demand can use any slot $s_2 \in \{s_1+1, \dots, s_3-1\}$. The following valid inequalities capture this fact:
\begin{align}
  & u_{des_1} + \sum_{d'\in D\setminus \{d\}} u_{d'es_2} + u_{des_3} \leq 2&
    \begin{tabular}{r}
      $\forall d \in D$,
      $\forall s_1,s_2,s_3 \in \{1, \dots, \bar s\}$, $s_1 < s_2 < s_3$.
    \end{tabular} \label{vi:nonOverBySum:1}
\end{align}
As the number of inequalities in this family may be too large, we take $s_3 = s_2 + 1$, thus allowing $s_1$ and $s_2$ to be in the ranges $[1, \bar s-2]$ and $[s_1+1, \bar s -1]$, respectively.

Let $D'\subseteq D$ be a minimal set of demands that do not fit in one arc, i.e., a set $D'$ such that $\sum_{d \in D'} v(d) > \bar s$ and for every $d'\in D'$, $\sum_{d \in D'\backslash\{d'\}} v(d) \leq \bar s$. Hence, for each arc $e\in E$, at least one demand from $D'$ cannot use $e$. Assuming each demand $d\in D'$ uses exactly $v(d)$ slots, this fact is captured by the optimality cuts
\begin{align}
  & \sum_{s' \in S} u_{des'} \leq v(d) \sum_{d' \in D'\backslash\{d\}} \left(v(d') - \sum_{s' \in S} u_{d'es'} \right)&
    \begin{tabular}{r}
      $d \in D'$, $\forall e \in E$.
    \end{tabular} \label{vi:KDemandsDoNotExceedS:1}
\end{align}
If we assume the constraints \refeq{DSL_BF:2}-\refeq{DSL_BF:7} and the family \emph{exactlyVdFromV}, then the inequalities \refeq{vi:nonOverBySum:1} do not imply \refeq{vi:KDemandsDoNotExceedS:1}, nor vice versa. We can consider an alternative optimality cut by summing over all slots used by the demands in $D'$ and forcing them to be less than or equal to the sum of their volumes minus the smallest one, namely
\begin{align}
  & \sum_{s \in S}\sum_{d \in D'} u_{des} \leq \sum_{d \in D'} v(d) - \min_{d \in D'}~v(d)&
    \begin{tabular}{r}
      $\forall e \in E$.
    \end{tabular} \label{vi:KDemandsDoNotExceedSBySum:1}
\end{align}
These two families of optimality cuts are not implied by each other. In our computational experiments, we separate the case $|D'|=2$, which attains a better performance.

Given a demand $d_1$ and an arc $e$, if a slot $s \le v(d_1)$ is used by another demand $d_2$, then $d_1$ cannot use any of the first $s$ slots. Moreover, $d_1$ cannot use any of the first $v(d_2)$ slots. We can express this fact individually for each of these first slots (and we call \emph{posFitLow2DBySlots} this familiy of inequalities), or we can express this fact by considering all these slots together in the inequality
\begin{align}
  & \sum_{s' = 1}^{s_1} u_{d_1es'} \leq s_2 (1 - u_{d_2es})&
    \begin{tabular}{r}
      $\forall e \in E$, $\forall d_1, d_2 \in D$, $d_1 \neq d_2$,
      $\forall s \in \{1, \dots, v(d_1)\}$,
    \end{tabular} \label{vi:positionalFittingLower:1}
\end{align}
where $s_1 = \max\{s, v(d_2)\}$ and $s_2 = \min\{v(d_1), s_1\}$, and we assume $d_1$ uses exactly $v(d_1)$ slots (so \refeq{vi:positionalFittingLower:1} is an optimality cut). An alternative version of this inequality is obtained by summing over every demand in $D\backslash\{d_1\}$ and relaxing the upper bound of the sum, by taking $s_1 = \max\{s, \min_{d'\in D\setminus\{d_1\}}(v(d'))\}$.

Let $d_1, d_2, d_3 \in D$ and $s_1,s_2,s_3 \in S$ be three different demands and slots. Then the inequality
\begin{align}
  & u_{d_1es_1} + u_{d_2es_2} + u_{d_3es_3} \leq 2&
    \begin{tabular}{r}
      $\forall e\in E$, 
      $v(d_1) \le s_1 < s_2 < s_3 \le \bar s-v(d_3)+1$,
      $s_3-s_1 \leq v(d_2)$,
    \end{tabular} \label{vi:posFit3DBySlots:1}
\end{align}
is valid for $RSA(G,D,\bar s)$. Variations of this inequality can be obtained by replacing $d_1$ and $d_3$ by the sum of all demands in $D\backslash\{d_2\}$.

For $s\in S$, define $D^{\geq}_s = \{d \in D: v(d) \geq s\}$ to be the set of demands with volumes greater than or equal to $s$, and define $D^< = D \backslash D^{\geq}$ to be the remaining demands. We can consider a variation of the inequalities \emph{posFitLow2DBySlots} by summing on $s_1$ the demands in $D^{\geq}_{s_2}$ and  summing on $s_2$ the demands in $D^<_{s_2}$, thus getting the inequality
\begin{align}
  & \sum_{d'\in D^{\geq}_{s_2}}u_{d'es_1} + \sum_{d' \in D^<_{s_2}}u_{d'es_2} \leq 1&
    \begin{tabular}{r}
      $\forall e\in E$,  $s_1 \in \{1, \dots, s_2-1\}$,
      $s_2 \in \{2, \dots, \max_{d \in D} v(d)\}$.
    \end{tabular} \label{vi:posFitLow2DBySlots2SetsOfDs:1}
\end{align}

Finally, for $e\in E$ and $d\in D$ with $v(d) > 1$, if any other demand uses a slot $s_2 \in [v(d)+1, 2v(d)]$ and $d$ uses a slot $s_1 < s_2$, then $d$ must use the central slots between $1$ and $s_2-1$. This is captured by
\begin{align}
  & \sum_{s' \in \gamma} u_{des'} \geq |\gamma| \Big(\sum_{d'\in D\setminus \{d\}} u_{d'es_2} + u_{des_1} - 1\Big)&
  \begin{tabular}{r}
    $\forall d \in D$, $e \in E$,
  \end{tabular}\label{vi:centralSsBetweenLowBounds:1}
\end{align}
where $s_2 \in \{v(d)+1, \dots, 2v(d)\}$, $s_1 < s_2$, and $\gamma = \{s_2-v(d), \dots, v(d)\}$. We can add a third demand using another slot as lower bound, thus getting
\begin{align}
  & \sum_{s' \in \gamma} u_{des'} \geq |\gamma| \Big(\sum_{d'\in D\setminus \{d\}} u_{d'es_1} + \sum_{d'\in D\setminus \{d\}} u_{d'es_3} + u_{des_2} - 2\Big)&
  \begin{tabular}{r}
    $\forall d \in D$, $e \in E$,
  \end{tabular} \label{vi:centralSsBetweenDs:1}
\end{align}
where $s_1 \in \{1, \dots, \bar s-2v(d)\}$, $s_3 \in \{s_1+ v(d)+1, \dots, s_1+2v(d)\}$, $s_1< s_2 < s_3$, and $\gamma = \{s_3-v(d), \dots, s_1+v(d)\}$.

\section{Separation procedures}\label{sepProcAndFamSel}
    We have implemented a branch-and-cut procedure for RSA in order to evaluate the contribution of these families of valid inequalities, valid equations, and optimality cuts within a cutting plane environment.

Each family is separated with an ad hoc separation procedure. Due to space limitations, we omit the details of their implementations. Their temporal complexities are polynomial in almost all cases, being $\mathcal{O}((|D|^{k} |E| \bar s^r)$ with $k, r \leq 3$ the worst. Exponential procedures that need complex structures, pre-compute a large set of them and take a random subset each time. As the effectiveness of each cut depends on its family, to be able to compare their effectiveness, we define a parameter $\epsilon$ for each separation procedure in such a way that a violated inequality is added as a cut if the absolute value of the difference between the left-hand side and the right-hand side is at least $\epsilon$. The higher the value of this parameter, the fewer cuts added. This parameter is calibrated for each family and fixed at its best value for the final experiments.

We also evaluated different strategies for managing the separation procedures. In some of them, we employ an \emph{effectiveness coefficient} for each procedure, defined as the number of generated cuts divided by the number of calls to the procedure. When the $\epsilon$ is calibrated it gives us a way to compare the behavior of the families. They also contemplate the variation of pre-sorting the list of procedures according to previous experiments.

\begin{itemize}
    \item \texttt{Brute Force} [BRUTE FORCE]:
    Execute all separation procedures.
    
    \item \texttt{Random} [RND]:
    Shuffle the list of procedures and iterate through it until at least one cut from $h$ different families is found, or until the end of the list is reached.
    
    \item \texttt{Most Effective} [EFF]:
    Sort the list of procedures by their effectiveness coefficient, and iterate through this list as in the \texttt{Random} strategy.
    
    \item \texttt{Most Effective With Random} [EFF\&RND]:
    Iterate over the sorted list of procedures as in \texttt{Most Effective}, but randomly call one of the non-called procedures with a predefined probability.
    
    \item \texttt{Weighted Selection} [WEIGHTED]:
    Iterate over the sorted list of procedures, and execute each procedure with a probability calculated as a function of its effectiveness coefficient.
\end{itemize}

\section{Computational results}\label{results}
   We now present our computational experience. The implementation was performed within the Cplex 12.10 environment, and the experiments were carried out on a computer with an Intel(R) Xeon(TM) 2.80GHz CPU with 4 GB of RAM memory. Both for the branch-and-cut and for the branch-and-bound we turned off all Cplex primal heuristics and pre-solving features.

The instance benchmark was generated using a script based on the literature, which is available at \cite{instances_generator}. Twenty real topologies were used, with $|V| \in \{6,\dots,43\}$ and $|E| \in \{9,\dots,176\}$. The number $\bar s$ of available slots depends on the instance, and it ranges from 5 to 200. The demands are randomly generated with uniform distribution, with volumes ranging from 1 to 124 slots. Every experiment is executed at least twice (three times for the parameters calibration) and the best result is selected.

To compare the results we define a coefficient $\tau$ for each run as follows. Let $t$ be the running time in minutes, let $g\in [0,1]$ be the optimality gap, and let $p=t/4$ be a penalty term. The coefficient $\tau$ is defined by
\begin{displaymath}
\tau = \left\{ \begin{array}{cl}
 t & \hbox{if the instance is solved within the time limit,} \\
 t + p + g*p & \hbox{if the instance is not solved within the time limit but a feasible solution is found,} \\
 t + 4p & \hbox{if no feasible solution is found within the time limit.}
 \end{array}
 \right.
\end{displaymath}
In this way, we penalize the lack of certainty when the time limit is reached, with a stronger penalty if no feasible solution was found. The lower the value of $\tau$, the better the result. The total coefficient $\tau$ of multiple runs is calculated as the sum of the best coefficient for each particular run.

\subsection{Selecting the most effective families of valid inequalities, valid equations, and optimality cuts}

    In order to calibrate the parameter $\epsilon$ for each procedure, we selected a subgroup of 22 small- and medium-sized instances. The execution time was limited to 4 minutes. For each family we experimented with $\epsilon \in \{ 0, 0.1, \dots, 2 \}$. For each such value of $\epsilon$ we calculated the coefficient $\tau$ for the best result of each instance, and as mentioned before we report the sum of all these coefficients (resulting in one coefficient for each combination). Table \refeq{table:eps_selection:1} shows these results for the best-performing families. It is interesting to note that the best results were obtained with small values of $\epsilon$, suggesting that these cuts are indeed effective.
    
    \begin{table}[ht]
        \begin{center}
            \begin{footnotesize}
                \begin{tabular}{c||cccccccccccccc}
                    Family $\backslash$ Eps.&0.0&0.1&0.2&0.3&0.4&0.5&0.6&0.7&0.8&0.9&1.0&1.1\\
                    \hline
                    \hline
                    contiguityIneqs&6.99&9.40&13.14&8.98&15.27&18.48&23.60&36.68&30.75&31.36&36.50&42.78\\
                    \hline
                    notBranchFromSrc&27.90&28.02&24.22&26.57&31.09&25.60&24.44&31.01&19.60&29.71&29.67&30.04 \\
                    \hline
                    farSlotsOff&26.43&25.87&29.06&23.15&23.60&23.98&29.95&27.39&31.22&23.81&26.32&29.00\\
                    \hline
                    contiguityEqs&25.57&34.98&26.62&23.85&31.35&30.97&25.61&37.81&39.13&23.89&35.87&42.17\\
                \end{tabular}
            \end{footnotesize}
        \end{center}
        \caption{Performance coefficient $\tau$ for the best-perfoming separation procedures.}
        \label{table:eps_selection:1}
    \end{table}
    
    Table \refeq{table:families_comparison:1} shows additional results of the some of the best families, reporting the sum of the best performance coefficients $\tau$ and the number of generated cuts for the best and worst values of $\epsilon$, respectively, and the mean and median values of the sum of $\tau$ in our experiments. The addition of almost every single separation procedure, with a proper parameter tuning, beats the simple branch-and-bound procedure, which attains a sum of the performance coefficient of $45.27$. This also holds for many poorly-calibrated values of $\epsilon$.
    
    \begin{table}[ht]
        \begin{center}
            \begin{footnotesize}
                \begin{tabular}{c||ccc|ccc|cc}
                    &\multicolumn{3}{c|}{Best Combination}&\multicolumn{3}{c|}{Worst Combination}&\multicolumn{2}{c}{Coefficient $\tau$} \\
                    Family &$\tau$&$\epsilon$&Cuts&$\tau$&$\epsilon$&Cuts&Mean&Median \\
                    \hline
                    \hline
                    contiguityIneqs&6.99&0.0&16499&45.51&4.0&0&32.80&39.46 \\
                    symmContiguityIneqs&8.25&0.1&14206&47.77&1.5&978&28.70&33.02 \\
                    symmetricalBFcontiguity&18.02&0.3&4427&34.29&1.9&3163&23.49&22.71 \\
                    notBranchFromSrc&19.60&0.8&4903&31.61&1.3&4094&27.00&27.23 \\
                    farSlotsOff&23.15&0.3&5349&47.35&1.4&2766&30.00&28.74 \\
                    ppalSlotsToDst&23.28&0.6&858&45.46&4.0&0&31.07&30.32 \\
                \end{tabular}
            \end{footnotesize}
        \end{center}
        \caption{Best, worst, median, and mean results for the best-performing procedures.}
        \label{table:families_comparison:1}
    \end{table}
    
\subsection{Comparison of separation strategies}

    We now report our experiments in order to evaluate the performance of the different separation strategies proposed in Section~5. 
    In order to calibrate the parameter $h$ (which limits the amount of different procedures to add cuts from) we used the same subgroup of 22 instances from the previous sub-section. We experimented with $h \in [5, 10, 15, 20, 25, 30]$. For each strategy and value of $h$, we considered the sum of the performance coefficient $\tau$ over all instances as a proxy for the overall performance. Once the best value of $h$ for each strategy was found, we experimented with the 94 instances with a timeout of 10 minutes. We eliminated from the results the instances for which all procedures stopped with a timeout, thus keeping 78 instances. Our cuts are added to the generic branch-and-bound implemented by cplex without pre-solve and primal heuristic, so the comparison is against this configuration, to show that our cuts are effective, and against the cplex branch-and-cut to prove that they are better than the generic ones. Table \refeq{table:strategies_comparison:1} reports these results, showing that all these strategies beat the two procedures implemented by Cplex. For the solved instances we improved all solution times with every strategy, whereas the best strategy was able to solve a third more instances than Cplex.
    
    \begin{table}[ht]
        \begin{center}
            \begin{footnotesize}
                \begin{tabular}{l||c|c|c|c|c|c|c|c||c|c}
                    &BRUTE FORCE&\multicolumn{2}{c|}{EFF\&RND Pre-sorted}&\multicolumn{2}{c|}{EFF Pre-sorted}&RND&\multicolumn{2}{c||}{WEIGHTED}&\multicolumn{2}{c}{CPLEX}\\
                    &h=38&h=10&h=5&h=10&h=5&h=20&h=15&h=25&B\&C&B\&B\\
                    \hline
                    \hline
                    $\tau$&585.41&559.29&503.63&576.67&380.63&483.51&500.16&498.27&792.60&1035.81 \\
                    \hline
                    Timeouts&25&24&21&25&13&19&21&19&35&46
                \end{tabular}
            \end{footnotesize}
        \end{center}
        \caption{Performance of the separation strategies for the 78 instances with no timeouts.}
        \label{table:strategies_comparison:1}
    \end{table}

\section{Conclusions}\label{conclusions}
       In this paper we explored valid inequalities, valid equations, and optimality cuts within a branch-and-cut environment for RSA. Since a polyhedral study of $RSA(G,D,\bar s)$ seems out of reach with the current knowledge, we settled at showing non-implications between these families. The computational experiments performed with the branch-and-cut procedure show that some of these families are quite effective, resulting in improvements over a simple branch-and-bound procedure and over the generic branch-and-cut procedure implemented by Cplex.
    
    Following the line of research taken, it would be interesting to explore the polytope $RSA(G,D,\bar s)$, in order to gain theoretical insights on the strength of the valid inequalities presented in this work. From a computational point of view, the separation procedures used in our implementation may be improved, e.g., by more effective heuristics or a better selection of which families to separate. As the cplex with pre-solve and heuristics turned on still beats our branch-and-cut the next step in the development of a fully-fledged integer-programming-based algorithm for RSA is the development of primal heuristics (both initial and rounding heuristics). We are currently working in this direction. We are currently performing experiments with larger time limits, in order to assess whether the behavior reported in this work carries on when the time limit is extended. Likewise we also ask ourselves whether it is possible to formulate the RSA without symmetries. 

\noindent\textbf{Acknowledgment.} We would like to thank the anonymous reviewers of this paper for their valuable comments and suggestions.

\end{document}